\documentclass[%
 reprint,
superscriptaddress,
 amsmath,amssymb,
 aps,
]{revtex4-2}

\usepackage{graphicx}
\usepackage{dcolumn}
\usepackage{bm}
\usepackage[utf8]{inputenc}
\usepackage[T1]{fontenc}
\usepackage{mathptmx}
\usepackage{etoolbox}
\usepackage{xcolor}
\usepackage{appendix}
\usepackage{chngcntr}

\begin{document}

\preprint{APS/123-QED}
 \title{Principles of Client Enrichment in Multicomponent Biomolecular Condensates}

 \begin{abstract}
Biomolecular condensates are commonly organized by a small number of \emph{scaffold} molecules that drive phase separation together with \emph{client} molecules that do not condense on their own but become selectively recruited into the dense phase. A central open question is how client recruitment feeds back on scaffold interactions to determine condensate composition. Here we address this problem in a reconstituted focal adhesion system composed of focal adhesion kinase (FAK) and phosphorylated p130Cas (Cas) as scaffolds and the adaptor protein paxillin (PXN) as a client. We show that both FAK phosphorylation and PXN recruitment produce a common compositional response in which FAK becomes enriched while Cas is depleted within the condensate. To interpret these observations, we develop two complementary theoretical descriptions. First, within a two-component Flory–Huggins framework, we show that phosphorylation can be captured by either strengthening heterotypic FAK–Cas interactions or increasing the effective number of interaction-relevant segments on FAK, both of which bias partitioning toward FAK-rich condensates. Second, we introduce a minimal three-component Flory–Huggins theory without an explicit solvent and map it onto an effective two-component description, demonstrating that client recruitment renormalizes homotypic and heterotypic scaffold interactions. Analytical predictions for the location of the critical point are tested in reconstituted multicomponent systems through PXN addition, showing that client recruitment alone tunes proximity to criticality and reshapes condensate composition. Together, our results reveal distinct yet convergent physical routes by which post-translational modification and client recruitment control scaffold composition in multicomponent condensates.
\end{abstract}

\author{Aishani Ghosal}
\affiliation{Department of Physics and Center for Biomolecular Condensates,
Washington University in St. Louis, St. Louis, MO 63130, USA}
\affiliation{School of Chemical Sciences, National Institute of Science Education and Research,  Bhubaneswar, Odisha 752050, India}

\author{Nicholas E. Lea}
\affiliation{Department of Biology,
Massachusetts Institute of Technology, Cambridge, MA, USA}

\author{Lindsay B. Case}
\email{lcase@mit.edu}
\affiliation{Department of Biology,
Massachusetts Institute of Technology, Cambridge, MA, USA}

\author{Trevor GrandPre}
\email{trevorg@wustl.edu}
\affiliation{Department of Physics and Center for Biomolecular Condensates,
Washington University in St. Louis, St. Louis, MO 63130, USA}
\affiliation{National Institute for Theory and Mathematics in Biology,
Northwestern University and The University of Chicago, Chicago, IL, USA}

\date{\today}
                       
\maketitle
\section{Introduction}

Biomolecular condensates are intracellular compartments that lack a surrounding membrane and may form via phase separation~\cite{harmon2017intrinsically,li2012phase,gomes2019molecular,shin2017liquid,zwicker2025physics,jacobs2023theory,lyon2021framework, mitrea2022modulating, pappu2023phase}. Typically, condensates comprise many molecular components, with some acting as scaffolds—the core molecules that drive and stabilize condensate formation— and others acting as clients (or ligands), which are recruited into the condensate but are not required for its assembly~\cite{banani2017biomolecular,banani2016compositional,wheeler2016distinct}. Although clients are not necessary for condensate formation, their enrichment can nonetheless modify condensate composition, emergent properties, and function. Despite significant progress on both fundamental and applied fronts, an open question is how client recruitment feeds back onto scaffold interactions and thereby reshapes condensate composition and partitioning.

The client-scaffold framework can be broadly applied to understand a wide range of biomolecular condensates in diverse biological contexts. In the C. elegans germline, P granules are biomolecular condensates containing more than 40 components and are essential for germline development and fertility. Their organization can be described in terms of scaffolds and clients, with scaffold proteins such as PGL family members (PGL-1 and PGL-3), MEG-3, and GLH helicases (GLH-1–4) contributing to condensate formation, and client molecules including Argonaute proteins (PRG-1, WAGO-1, ALG-3, and CSR-1), mRNAs, and other RNA-binding proteins becoming selectively enriched~\cite{brangwynne2009germline,updike2011p,schisa2001rna, updike2010p, saha2016polar}. In yeast, cytoplasmic processing bodies (P bodies) regulate RNA metabolism. RNA is a critical scaffold molecule for P bodies, in addition to redundant mutlivalent scaffold proteins (Dhh1, Edc3, Xrn1, Dcp1, and Pat1)~\cite{sheth2003decapping, decker2012p}. P bodies also contain numerous client-like proteins such as Not2, CCR4, and Pop2~\cite{sheth2003decapping}. Focal adhesions are dynamic assemblies at the cell–extracellular matrix interface of animal cells that contain hundreds of distinct proteins and play a central role in cell adhesion, migration, and mechanotransduction~\cite{case2015integration, chastney2021integrin, pang2023targeting}. Growing evidence suggests that focal adhesions are condensates that can form through phase separation, and their composition can be described in terms of scaffolds and clients. The multivalent signaling proteins FAK and p130Cas undergo phase separation and are required for focal adhesion formation, while other adhesion-associated proteins, including vinculin and the Arp2/3 complex, become selectively enriched and modulate focal adhesion stability~\cite{swaminathan2016fak, Case2022synergistic, hoffmann2014symmetric, le2022new, liang2024paxillin, martin2025conformation, kumari2024focal, hu2014fak}. 

Although clients are not required for condensate formation, client molecules can influence scaffold interactions to remodel condensates. The influence of client molecules on condensates has been most clearly demonstrated in reconstituted in vitro systems with a single scaffold component. Client enrichment can destabilize condensates by sequestering scaffold interaction sites or altering network connectivity, highlighting that client recruitment can either stabilize or dissolve condensates depending on molecular context~\cite{dao2018ubiquitin,ahmed2025client}. Client molecules can also remodel condensates in subtler ways, including stabilizing size-limited microphases or selectively destabilizing specific interaction networks without eliminating phase separation altogether~\cite{shinn2026nuclear}. Together, these experimental observations demonstrate that clients can modulate condensate formation, composition, and size. However, how clients impact multicomponent condensates with multiple scaffolds has not been extensively studied. 

In living cells, condensates are not static: their formation, dissolution, and material properties are regulated by active, energy-consuming processes that operate far from equilibrium. Cells can modulate condensate properties, including composition, material state, and viscosity, through posttranslational modifications such as phosphorylation~\cite{kasahara2018phosphorylation, kim2019phospho}. In focal adhesion condensates, phosphorylation of tyrosine residues within the scaffold p130Cas creates multivalent SH2 domain binding sites that enhance phase separation and condensate formation. Preventing this modification via tyrosine-to-phenylalanine mutation reduces phase separation with adaptor proteins in vitro, leading to fewer condensates in vitro and decreases focal adhesion formation in cells~\cite{Case2022synergistic}. 

From a physical perspective, phosphorylation tunes both the strength and number of binding interactions among condensate components, thereby modulating multivalent attractions and phase behavior~\cite{liu2024crosstalk, hofweber2019friend}. Importantly, these effects are context-dependent: in some systems, such as Fused in Sarcoma (FUS), phosphorylation introduces electrostatic repulsion or disrupts interaction motifs, weakening connectivity and suppressing condensation~\cite{monahan2017phosphorylation}. More generally, kinase-mediated phosphorylation acts as a nonequilibrium control mechanism that reshapes interaction networks in space and time and can, in some cases, dynamically alter the functional roles of individual molecules, for example shifting components between client-like and scaffold-like behavior. Understanding how such local biochemical modifications propagate through multicomponent condensates to control their collective composition and function remains a central challenge.


\begin{figure}[h]
\centering
\includegraphics[width=0.4\textwidth]{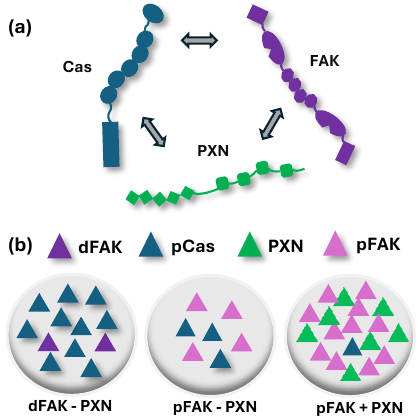}
\caption{\label{fig:0} 
Schematic illustrating the effects of phosphorylation and client proteins on condensate composition and concentration.
(a) Phosphorylated Cas (pCas) and FAK act as scaffolds, while PXN serves as a client; weak multivalent interactions are indicated by double-sided arrows.
(b) Scaffolds and the client are represented by  triangular symbols of different colors. 
Grey circles represent condensates composed of (left) dFAK-pCas; (middle) pFAK-pCas; (right) pFAK-pCas-PXN.
It shows how the condensate composition changes upon FAK phosphorylation (middle) and FAK phosphorylation combined with client enrichment (right), compared to their absence (left).}
\end{figure}

A growing body of theoretical and computational work has clarified how scaffold–client interactions regulate condensate formation, stability, and organization~\cite{alberti2025current, fare2021higher, ditlev2018s}. Thermodynamic theory based on polyphasic linkage, together with coarse-grained simulations, shows that clients need not act as inert cargo: depending on their binding mode, affinity, and spatial architecture, they can either promote phase
separation by introducing additional crosslinks or suppress it by sequestering scaffold interaction sites and weakening scaffold–scaffold connectivity~\cite{espinosa2020liquid, ruff2021ligand, wyman1980ligand, galagedera2023polyubiquitin, valentino2024phase}. Complementary multi-scale simulations and analytical polymer-physics models demonstrate that scaffold–client specificity can emerge from sequence-encoded multivalent interactions among intrinsically disordered regions, even in the absence of ordered binding sites~\cite{gaurav2025multi,wessen2025sequence}. Building on this, post-translational modifications have been shown to tune client–scaffold interactions and client partitioning, thereby regulating reaction rates and efficiency without substantially altering the scaffold-rich phase~\cite{laha2024chemical}. Separately, theory shows that scaffold–client interactions regulate condensate size, kinetics, internal organization, and interfacial exchange through client concentration and valency~\cite{sanchez2021size,sanchez2021valency,rana2024interfacial, w7g3-6rsd}. While existing theory establishes that client enrichment and phosphorylation can regulate condensate properties, how these processes quantitatively determine
condensate scaffold composition in multicomponent systems remains largely unexplored.

In this work, we investigate how client recruitment and phosphorylation regulate the composition of multicomponent biomolecular condensates. We focus on a model system consisting of two scaffold proteins, p130Cas (hereafter Cas) and focal adhesion kinase (FAK), together with the client protein paxillin (PXN) (Fig.~\ref{fig:0}A). We've previously shown that both Cas and FAK can independently drive phase separation, and Cas, FAK and PXN can all directly interact with each other and colocalize within multicomponent condensates~\cite{Case2022synergistic}. Using reconstituted in vitro experiments, we show that FAK phosphorylation and PXN recruitment drive selective enrichment of FAK and concomitant dilution of Cas within the condensed phase (schematized in Fig.~\ref{fig:0}B). To understand the thermodynamic origin of these compositional changes, we use Flory–Huggins theory~\cite{flory1953principles, huggins1941solutions}. Our analysis reveals that phosphorylation can be modeled as increasing the heterotypic interactions and potentially reducing the mixing entropy of FAK, while client recruitment can be modeled as effectively weakening Cas–Cas homotypic interactions and enhancing FAK–FAK homotypic interactions. We further compute critical concentrations and heterotypic interaction
thresholds for phase separation in the presence of a client, demonstrating that introducing a client capable of both self-interaction and scaffold binding effectively renormalizes scaffold–scaffold interactions, lowers the critical interaction strength required condensation, and reshapes condensate composition.

\begin{figure*}
\centering
\includegraphics[width=0.8\textwidth]{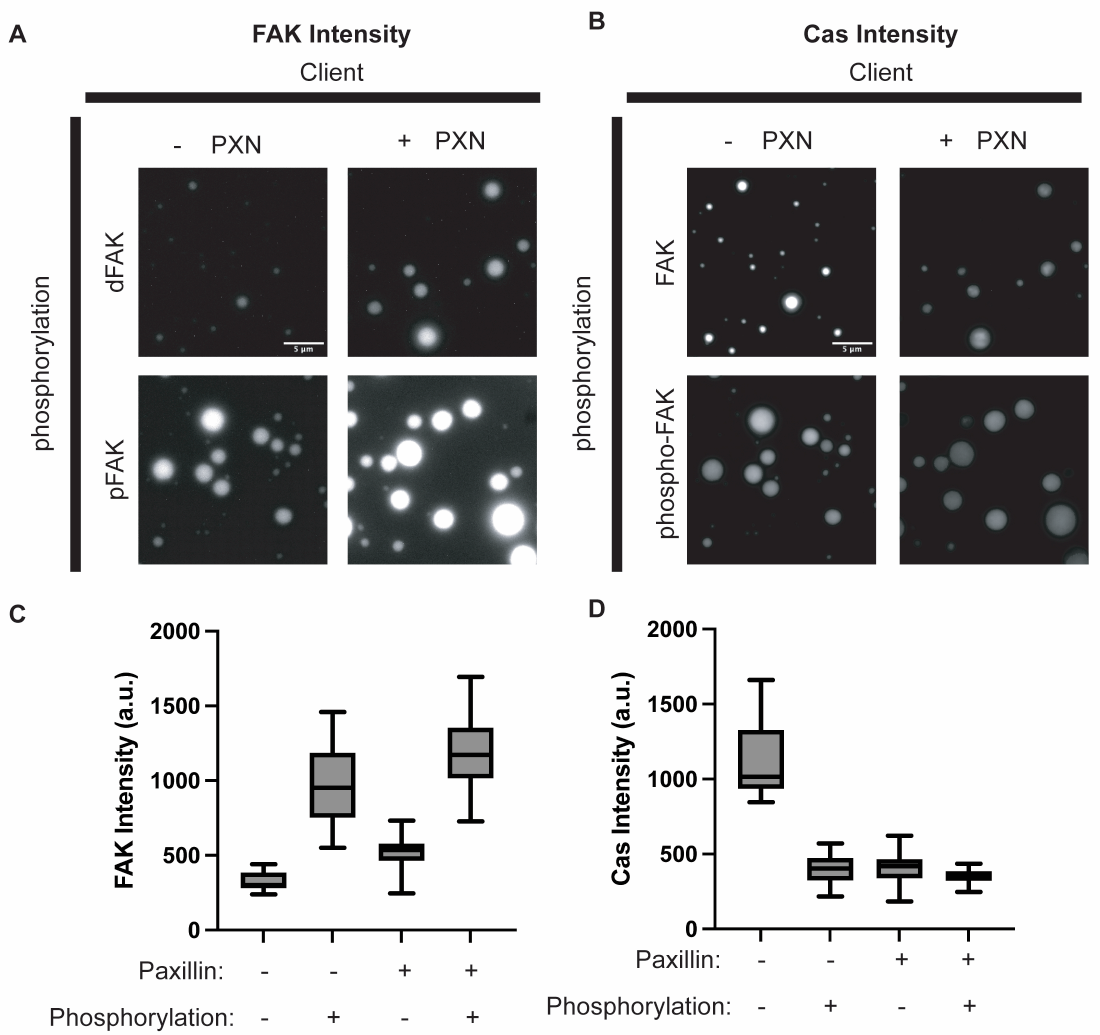}
\caption{\label{fig:1}  Fluorescence microscopy images of focal adhesion condensates formed via phase separation.
Condensates were assembled using four distinct protein compositions by combining all components at a concentration of 1~$\mu$M in buffer containing 50~mM HEPES (pH~7.5), 50~mM NaCl, 1~mM DTT, and 1~mg/mL BSA. Images were acquired on an epifluorescence microscope 20~minutes after mixing to allow the system to reach equilibrium.
The four compositions tested were: dFAK$-$PXN (dephosphorylated FAK, 15\% GFP-labeled; phosphorylated Cas, 5\% Alexa647-labeled; Nck; and N-WASP), dFAK$+$PXN (same as dFAK$-$PXN with the addition of PXN), pFAK$-$PXN (phosphorylated FAK, 15\% GFP-labeled; phosphorylated Cas, 5\% Alexa647-labeled; Nck; and N-WASP), and pFAK$+$PXN (same as pFAK$-$PXN with the addition of PXN).
(A) GFP channel images used to quantify FAK intensity within condensates.
(B) Alexa647 channel images used to visualize Cas intensity.
Panels A and B correspond to the same condensates. Scale bars are 5~$\mu$m.
(C,D) Quantification of fluorescence intensity within condensates. Images were thresholded and condensates segmented using ImageJ. Fluorescence intensities (arbitrary units, a.u.) of FAK (GFP) and Cas (Alexa647) were measured within condensates, and mean condensate intensities were calculated for each image. Experiments were repeated three times with five images collected per replicate. Data are displayed as box plots. Fluorescence intensities of FAK and Cas are not quantitatively comparable due to the use of different fluorophores and camera exposure settings.
}
\end{figure*}
The paper is organized as follows. In Sec.~\ref{Sec:exp}, we present experimental evidence using fluorescence imaging and intensity-profile analysis to measure relative scaffold dense phase concentrations under physiological perturbations, including phosphorlyation and client recruitment. In Secs.~\ref{Sec:phosphorylation}–\ref{Sec:criticalparams}, we develop a mean-field theoretical framework to interpret these experimental observations. The role of tyrosine phosphorylation of FAK in regulating condensate composition and phase demixing is examined in Sec.~\ref{Sec:phosphorylation}. Section~\ref{Sec:PXNaddition} analyzes how the presence of the client protein PXN reshapes condensate composition. The combined effects of phosphorylation and PXN recruitment are discussed in Sec.~\ref{Sec:botheffects}. In Sec.~\ref{Sec:criticalparams}, we show how PXN modifies the critical interaction parameters governing phase separation. In Sec.~\ref{exper}, we show results from reconstituted experiments which confirm our theory that PXN decrease the critical point of condensation. Finally, Sec.~\ref{Sec:Conclusion} summarizes our findings and discusses their broader implications.

\section{Result}\label{Sec:Result}
\subsection{in vitro experiments to measure condensate composition}\label{Sec:exp}

In this study, we use reconstituted focal adhesion condensates as a model system to explore the principles that regulate composition in multicomponent condensates. Using purified recombinant proteins, we previously showed that both Cas and FAK are scaffold molecules that can drive phase separation and condensate formation~\cite{Case2022synergistic}. Cas undergoes phase separation when mixed with the adaptor proteins Nck and N-WASP, due to multivalent interactions between these three components. For simplicity, we will refer to these multicomponent Cas-Nck-N-WASP mixtures as Cas. FAK undergoes homotypic phase separation driven by interactions between its folded domains. Although the protein PXN does not undergo homotypic phase separation at physiological concentrations ~\cite{Case2022synergistic, liang2024paxillin}, it contains an IDR that can weakly self-associate\cite{liang2024paxillin}. FAK, Cas, and PXN all interact directly with each other (Fig. 1a), and when all three components are combined, they colocalize in multicomponent condensates~\cite{Case2022synergistic}. While Cas and FAK are scaffolds of these multicomponent condensates, PXN is more client-like. 

Additionally, FAK can be phosphorylated on several residues, 
and FAK phosphorylation creates additional interactions between FAK and CAS through the Nck SH2 domain~\cite{goicoechea2002nck}. 
To determine the impact of phosphorylation and client recruitment on condensate composition, we reconstituted condensates with different combinations of proteins (Fig. 2). All proteins were purified as previously described~\cite{Case2022synergistic}. FAK purified from insect cells is phosphorylated. To compare phosphorylated and dephosphorylated FAK, we dephosphorylated FAK with the phosphatase PTP1 as previously described~\cite{lea2025kinase}. FAK is labeled with GFP and Cas is labeled with Alexa647. From fluorescence microscopy images, we quantified the FAK intensity inside condensates and the Cas intensity inside condensates. FAK phosphorylation increases FAK intensity and decreases Cas intensity inside condensates. Similarly, adding paxillin increases FAK intensity and decreases Cas intensity inside condensates. These experiments demonstrate that FAK phosphorylation and paxillin recruitment change the concentrations of both scaffolds (FAK and Cas) in the dense phase. The effects of phosphorylation and paxillin are also additive. To understand the thermodynamic origin of these compositional changes, we turned to Flory–Huggins theory.

\subsection{Theory of FAK phosphorylation}
\label{Sec:phosphorylation}
\begin{figure*}
\centering
\includegraphics[width=0.85\textwidth]{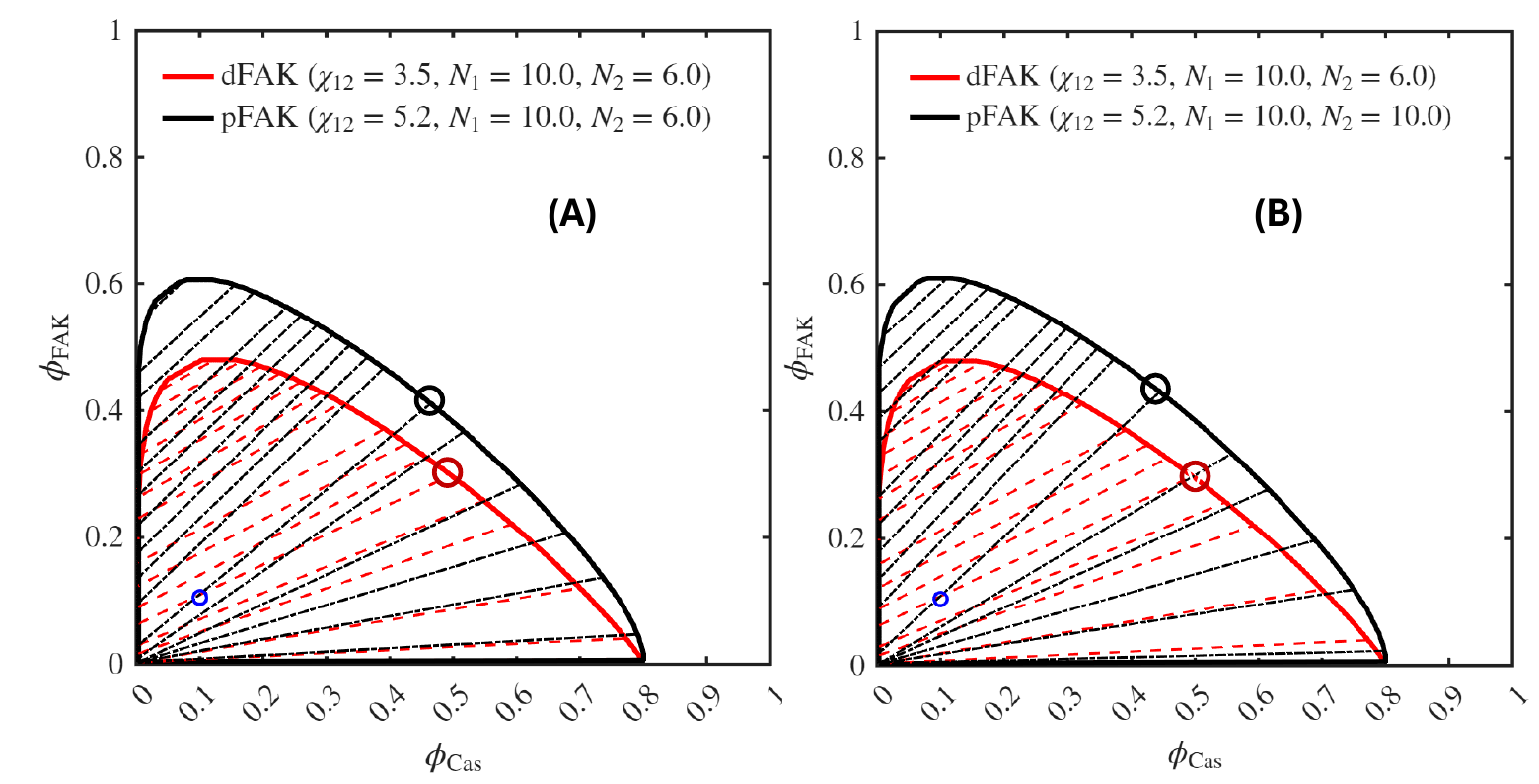}
\caption{\label{fig:2componentfull} 
The phase diagrams are constructed from the Flory–Huggins free energy (Eq.~\eqref{eq:FH3} with $M = 2$) using a convex-hull algorithm. 
Red dashed lines indicate tie lines connecting the dilute and dense phases prior to FAK phosphorylation, while black dash-dotted lines denote tie lines after phosphorylation.
The red and black binodals correspond to ternary systems composed of dFAK–Cas–solvent and pFAK–Cas–solvent, respectively. 
Phosphorylation of FAK is assumed to (A) enhance the FAK–Cas interaction ($\chi_{12}$) or (B) increase both the number of stickers on FAK ($N_2$) and  FAK–Cas interaction ($\chi_{12}$), while keeping the Cas–Cas ($\chi_{11}$) and FAK–FAK ($\chi_{22}$) interactions, as well as the number of stickers on Cas ($N_1$), constant. The parameters are shown in the legend, and the rest of the parameters used in the phase diagram construction are: $\chi_{11} = 2.8$, $\chi_{22} = 0.2$.
The blue circle denotes $\phi_{\mathrm{Cas}} = 0.075$ and $\phi_{\mathrm{FAK}} = 0.096$, while the red and black circles denote the dense-phase compositions before and after phosphorylation of FAK, respectively.
The circles show FAK phosphorylation is found to increase the FAK concentration while depleting Cas within the condensate; nevertheless, it renders the dense phase overall denser, as indicated by the larger area enclosed by the binodal. 
}
\end{figure*}
We model FAK, Cas, and PXN as sticker–spacer polymers and qualitatively compute the dilute- and dense-phase compositions from the Flory–Huggins (FH) free energy~\cite{flory1953principles,huggins1941solutions}, which is given by
\begin{widetext}
\begin{equation}
f = \sum_{i=1}^{M} \frac{\phi_i}{N_i}\ln \phi_i
+ \left(1-\sum_{i=1}^{M}\phi_i\right)\ln\!\left(1-\sum_{i=1}^{M}\phi_i\right)
-\frac{1}{2}\sum_{i,j=1}^{M}\chi_{ij}\phi_i\phi_j .
\label{eq:FH3}
\end{equation}
\end{widetext}
Equation~\eqref{eq:FH3} gives the Flory--Huggins free energy density for an incompressible multicomponent polymer solution, expressed in units of $k_BT$ per coarse-grained lattice volume. 
The first two terms describe the entropy of mixing of the solute components and the solvent, while the final term accounts for effective pairwise interactions between components, measured relative to their interactions with the solvent through the Flory interaction parameters $\chi_{ij}$~\cite{deviri2021physical, jacobs2023theory}. 

The free energy density $f$ is defined per lattice site, corresponding to a coarse-grained polymer segment, such that the $\phi_i$ are volume fractions normalized by the segment volume, $v=1$.
The parameter $N_i$ denotes the degree of polymerization of species $i$, measured in units of these coarse-grained segments.
Importantly, within Flory--Huggins theory $N_i$ counts the number of independently translating segments along a polymer chain and therefore controls the translational entropy of mixing.

Heterotypic scaffold--client interactions are captured separately through the corresponding interaction parameters $\chi_{ij}$. Positive values of the Flory--Huggins interaction parameters $\chi_{ij}$ correspond to net attractive interactions between components $i$ and $j$, relative to their interactions with the solvent. 
Interactions with identical indices ($i=j$) represent homotypic self-interactions, whereas interactions with distinct indices ($i\neq j$) represent heterotypic interactions between different components. To compute phase coexistence, we evaluate the mean-field Flory--Huggins free energy and determine the binodal curves using a convex-hull construction, which identifies the globally stable phase-separated states by enforcing equality of chemical potentials and osmotic pressure~\cite{mao2019phase,lee1992computer,wolff2011thermodynamic}. 

Since both FAK and Cas (representing Cas–Nck–N-WASP) undergo phase separation when individually mixed with solvent, we model each scaffold as having positive (attractive) homotypic interactions. 
Because the two proteins also contain binding sites for one another, their heterotypic interaction is likewise taken to be nonzero and attractive. 
Accordingly, we choose interaction parameters such that
\[
\chi_{11} > 0,\quad \chi_{22} > 0,\quad \chi_{12} > 0.
\]
We focus on the regime $\chi_{12}>\chi_{11}$, which favors associative phase separation to compute the resulting phase diagrams.

 In Fig.~\ref{fig:1}C and D, phosphorylation leads to enrichment of FAK and depletion of Cas within the condensate. Phosphorylation is thought to enhance heterotypic scaffold–scaffold interactions. In Fig.~\ref{fig:2componentfull}A, we illustrate the effect of increasing $\chi_{12}$ on the phase behavior of the FAK–Cas system. 
As $\chi_{12}$ increases, the binodal (red curve) expands, indicating an enlarged two-phase coexistence region and a denser condensate enriched in both FAK and Cas. The area enclosed by the binodal corresponds to compositions that phase separate into coexisting dense and dilute phases, whereas compositions outside the binodal remain homogeneously mixed. 
Tie lines, shown as dashed lines, connect the compositions of the coexisting phases and illustrate how scaffold concentrations partition between the dilute and condensed phases along each coexistence condition. Our theory shows that increasing heterotypic interactions, $\chi_{12}$, alters the slope of the tie lines, producing a compositional shift consistent with experiments in which Cas is less enriched and FAK is more enriched in the dense phase.

\begin{figure}
\centering
\includegraphics[width=0.45\textwidth]{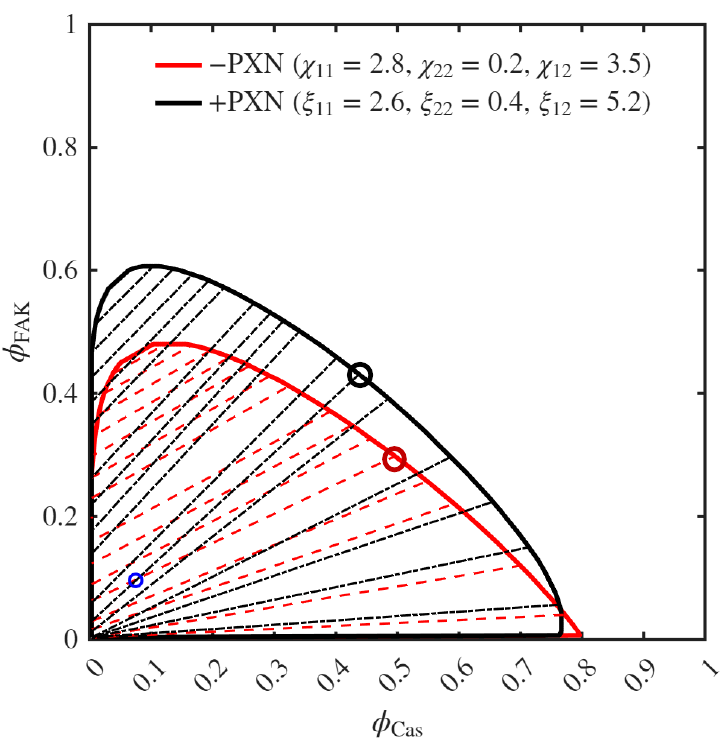}
\caption{\label{fig:2componentreduced}
Theoretical prediction for the effect of the client PXN on Cas–FAK phase separation in an effective two-component Flory–Huggins framework.
Phase diagrams are constructed using the convex hull method from the Flory-Huggins free energy (Eq.~\eqref{eq:FH3} for dFAK-pCas system and 
Eq.~\eqref{eq:FH3nosolvnewparams} for dFAK-pCas-PXN system), and plotted with the volume fractions of Cas (FAK), denoted by $\phi_{\text{Cas}}$ ($\phi_{\text{FAK}}$), along the horizontal (vertical) axes. 
Red solid binodal corresponds to dFAK–Cas mixture in the presence of an explicit solvent, whereas black solid binodal refers to the dFAK–Cas–PXN system without an additional solvent, with PXN effectively acting as the solvent. 
In dFAK–Cas mixture, interactions between the two scaffolds include both homotypic and heterotypic contributions, described by the Flory–Huggins parameters $\chi_{11}$, $\chi_{22}$, and $\chi_{12}$. 
Additionaly, dFAK–Cas-PXN mixture incorporates scaffold–client interactions, which modifies the scaffold–scaffold interactions and lead to effective homotypic and heterotypic interaction parameters for the scaffolds. 
The resulting effective interaction parameters are $\xi_{1} = \chi_{11} + \chi_{33} - 2\chi_{13}$, $\xi_{2} = \chi_{22} + \chi_{33} - 2\chi_{23}$, and 
$\xi_{3} = \chi_{12} + \chi_{33} - \chi_{13} - \chi_{23}$.
The parameters used to construct the phase diagrams are as follows: (Red) $\chi_{11} = 2.8$, $\chi_{22} = 0.2$, $\chi_{12} = 3.5$; $N_{1} = 10$, $N_{2} = 6$; (Black) $\chi_{11} = 2.8$, $\chi_{22} = 0.2$, $\chi_{33} = 1.0$, $\chi_{12} = 5.2$, $\chi_{13} = 0.6$ and $\chi_{23} = 0.4$, $N_{1} = 10$; $N_{2} = 10$, which lead to the effective interaction parameters $\xi_{1} = 2.6$, $\xi_{2} = 0.4$, and $\xi_{3} = 5.2$. 
The blue circle denotes $\phi_{\mathrm{Cas}} = 0.075$ and $\phi_{\mathrm{FAK}} = 0.096$, while the red and black circles denote the dense-phase compositions before and after the addition of PXN, respectively.}
\end{figure}

Phosphorylation of FAK may have effects beyond strengthening heterotypic interactions with Cas. Although it is not generally thought to substantially enhance homotypic FAK–FAK or Cas–Cas interactions, nor to dramatically alter solvent quality, phosphorylation could modify the entropy of mixing of FAK. To capture this effect within a Flory–Huggins framework, we redefine the coarse-graining length used to describe the scaffold: rather than counting monomeric units, we model FAK in terms of sticker–spacer modules and introduce an effective polymerization index, $N_2$, that increases upon phosphorylation. This renormalization reduces the translational entropy of FAK without changing its homotypic interaction parameter, thereby biasing partitioning toward the condensed phase. As a result, the system remains in the associative phase-separation regime, but the condensate becomes selectively enriched in FAK and depleted in Cas. 
In Fig.~\ref{fig:2componentfull}B, we demonstrate that simultaneously increasing the heterotypic FAK–Cas interaction and the effective sticker number of FAK reproduces this experimentally observed enrichment–depletion behavior.

Changing the effective chain length and heterotypic interaction strength alters the critical composition at which phase separation first becomes unstable. Within the Flory--Huggins framework, the critical point $(\phi_1^c,\phi_2^c)$ marks the composition at which the binodal and spinodal meet, and therefore provides a sensitive measure of how interactions bias condensate composition at onset. In our model, phosphorylation of FAK is represented by an increase in both the FAK-Cas heterotypic interaction strength $\chi_{12}$ and the effective degree of polymerization of FAK, $N_2$, while Cas properties are held fixed. These changes shift the critical point toward higher FAK volume fraction and lower Cas volume fraction, corresponding to an increase in $\phi_2^c$ and a decrease in $\phi_1^c$.   Consequently, at fixed interaction parameters, the phosphorylated system lies further from criticality than the dephosphorylated system, resulting in an expanded two-phase coexistence region without requiring changes to Cas-Cas homotypic interactions.

\subsection{Model of recruitment of PXN}\label{Sec:PXNaddition}

As shown in Fig.~\ref{fig:1}C and D, the addition of PXN to FAK–Cas condensates produces an effect qualitatively similar to phosphorylation: the condensed phase becomes enriched in FAK and depleted in Cas. To understand the physical origin of this compositional shift, we extend our theoretical description to a three-component Flory–Huggins (FH) model comprising FAK, Cas, and PXN. For analytical tractability, we consider a solvent-free formulation in which incompressibility is enforced explicitly among the three components.

We, therefore, use Eq.~\eqref{eq:FH3} with $M=3$, neglecting the explicit solvent entropy term. In this formulation, FAK and Cas act as scaffolds, while PXN is treated as a client that does not phase separate independently but engages in both homotypic and heterotypic interactions. The presence of PXN introduces additional interaction parameters $\chi_{33}$, $\chi_{13}$, and $\chi_{23}$, corresponding to PXN–PXN, Cas–PXN, and FAK–PXN interactions, respectively. This choice is motivated by experimental evidence that PXN binds directly to both FAK and Cas and can self-associate weakly under crowded conditions~\cite{Case2022synergistic, liang2024paxillin}. We further assume $\chi_{23}>\chi_{13}$ to reflect the stronger and more multivalent interactions of PXN with FAK compared to Cas, consistent with its preferential enrichment in FAK-rich condensates observed experimentally~\cite{Case2022synergistic}.

We can now quantify how PXN client--client and scaffold--client interactions reshape FAK--Cas condensation by explicitly eliminating the PXN volume fraction using the incompressibility constraint $\phi_{3}=1-\phi_{1}-\phi_{2}$. Substituting this relation into the three-component Flory--Huggins free energy yields an effective two-component description in which the presence of PXN renormalizes the scaffold--scaffold interactions. Terms that are linear in the volume fractions are omitted, as they do not affect phase equilibria or stability conditions.

\begin{widetext}
\begin{equation}
\begin{aligned}
f(\phi_{1},\phi_{2}) &= \frac{\phi_{1}}{N_{1}} \ln \phi_{1}
+ \frac{\phi_{2}}{N_{2}} \ln \phi_{2}
+ \frac{1}{N_{3}} (1-\phi_{1}-\phi_{2}) \ln (1-\phi_{1}-\phi_{2}) \\
&\quad -\frac{1}{2}\,\xi_{1}\phi_{1}^{2}
- \frac{1}{2}\,\xi_{2}\phi_{2}^{2}
- \xi_{3}\phi_{1}\phi_{2},
\label{eq:FH3nosolvnewparams}
\end{aligned}
\end{equation}
\end{widetext}
where $\xi_{1} = \chi_{11} + \chi_{33} - 2\chi_{13}$, $\xi_{2} = \chi_{22} + \chi_{33} - 2\chi_{23}$,
and  $\xi_{3} = \chi_{12} + \chi_{33} - \chi_{13} - \chi_{23}$.

In Fig.~\ref{fig:2componentreduced}, we illustrate the effect of PXN addition by comparing a reference system in which PXN is non-interacting ($\chi_{33}=\chi_{13}=\chi_{23}=0$) to a system in which PXN exhibits attractive homotypic and scaffold--client interactions ($\chi_{i3}>0$ for $i=1,2,3$). Introducing PXN renormalizes the effective scaffold interactions: the effective FAK--FAK interaction strength $\xi_{2}$ increases, while the effective Cas--Cas interaction strength $\xi_{1}$ decreases, even though the bare scaffold--scaffold interactions are unchanged. As a result, condensates become enriched in FAK and depleted in Cas at fixed heterotypic scaffold--scaffold coupling. In this framework, PXN does not act as an inert solvent; instead, its attractive interactions with both scaffolds lower the free-energy cost of concentrating FAK-rich phases while effectively diluting Cas self-association. This mapping makes explicit how client recruitment can reshape scaffold-driven condensates by renormalizing effective interaction strengths and biasing condensate composition, without requiring direct changes to the intrinsic homotypic interactions of the scaffolds themselves.

\subsection{FAK Phosphorylation and Recruitment of PXN}\label{Sec:botheffects}

We find that both FAK phosphorylation and PXN recruitment independently drive enrichment of FAK and depletion of Cas within the condensed phase. As shown in Fig.~\ref{fig:1}C and D, when phosphorylation and PXN addition are applied simultaneously, their effects are compounded: FAK becomes further enriched while Cas is further depleted from the condensate.

\begin{figure}
\includegraphics[width=0.45\textwidth]{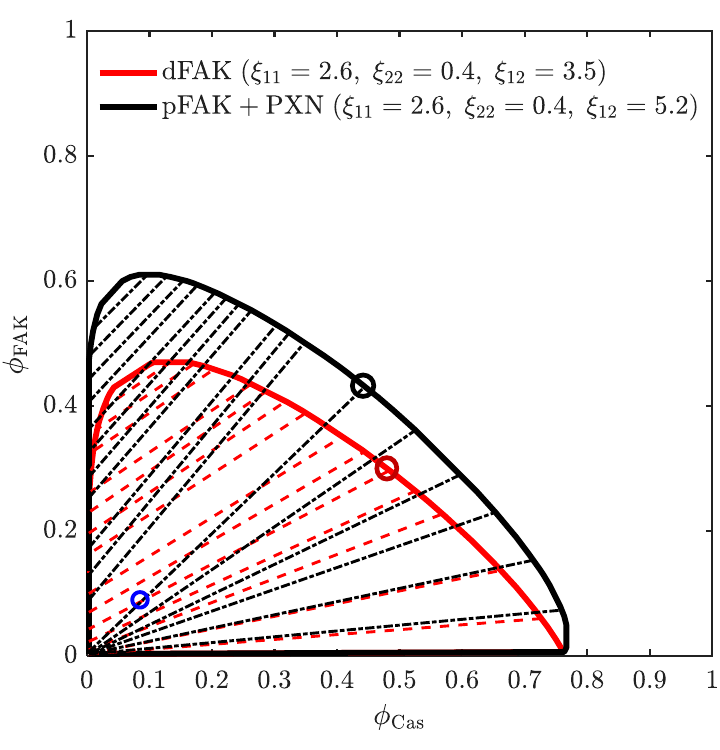}
\caption{\label{fig:PXN_phosphorylation_effect} 
Effect of phosphorylation and client (PXN) addition on the phase diagrams of a system consisting of two scaffolds (FAK and Cas) and one client (PXN), with no solvent. 
Phase diagrams are constructed using Flory-Huggins free energy expressed in Eq.~\eqref{eq:FH3} and Eq.~\eqref{eq:FH3nosolvnewparams}.
FAK phosphorylation effectively modifies the heterotypic scaffold–scaffold interactions ($\chi_{12}$) and FAK sticker numbers ($N_2$), while client addition also alters all effective homotypic and heterotypic interactions among the scaffolds.
The effective Flory–Huggins interaction parameters ($\xi_{1}$, $\xi_{2}$, and $\xi_3$) between the scaffolds are defined in terms of the scaffold–scaffold and scaffold–client interaction parameters, as described after Eq.~\eqref{eq:FH3nosolvnewparams}.
The actual and effective interaction parameters are as follows: 
(Red) $\chi_{11} = 2.8$, $\chi_{22} = 0.2$, $\chi_{33} = 1.0$, $\chi_{12} = 3.5$, $\chi_{13} = 0.6$ $\chi_{23} = 0.4$, $N_{1} = 10$, and $N_{2} = 6$, yielding effective interactions $\xi_{1} =2.6$, $\xi_{2} = 0.4$ and $\xi_{3} = 3.5$;
(Black) $\chi_{11} = 2.8$, $\chi_{22} = 0.2$, $\chi_{33} = 1.0$, $\chi_{12} = 5.2$, $\chi_{13} = 0.6$ $\chi_{23} = 0.4$, $N_{1} = 10$, and $N_{2} = 10$, yielding effective interactions $\xi_{1} = 2.6$, $\xi_{2} = 0.4$ and $\xi_{3} = 5.2$. 
The blue circle denotes $\phi_{\mathrm{Cas}} = 0.085$ and $\phi_{\mathrm{FAK}} = 0.09$, and red and black circles denote the dense phase compositions before and after the addition of PXN and phosphorylation of FAK, respectively. 
The dense-phase is enriched in FAK and depleted in Cas upon phosphorylation and PXN addition.}
\end{figure}

In Fig.~\ref{fig:PXN_phosphorylation_effect}, we demonstrate that this enhanced compositional bias can be captured by combining the parameter changes associated with each mechanism. Phosphorylation is modeled as an increase in heterotypic interactions, $\chi_{12}$, and an increase in effective degree of polymerization of FAK $N_2$, corresponding to a reduction in its mixing entropy, while PXN recruitment is modeled through renormalization of the effective scaffold--scaffold interactions. Using the parameter regimes established in Figs.~\ref{fig:2componentfull} and \ref{fig:2componentreduced}, we show that these combined effects are sufficient to reproduce the experimentally observed amplification of FAK enrichment and Cas depletion, without invoking additional changes to Cas intrinsic interactions or solvent quality.

In the following section, we examine how these combined effects alter the critical concentrations and interaction parameters governing phase separation.

\subsection{Effect of phosphorylation and clients on critical points}\label{Sec:criticalparams}

Critical conditions in Flory--Huggins models are classically obtained by identifying points at which the homogeneous free energy loses stability, corresponding to the simultaneous vanishing of the second and third derivatives of the free energy regarding composition, 
as established for binary polymer solutions~\cite{ rubinstein2003polymer, safran2018statistical,de2024exact, qian2022analytical}. Extensions of this framework to multicomponent polymer systems have been explored using analytical and numerical approaches in the polymer physics literature~\cite{deviri2021physical}, 
but closed-form expressions for critical parameters in three-component scaffold-client models have not, to our knowledge, been derived in this context. 

Here, we consider a three-component system composed of two scaffolds and one client, with free-energy density given by Eq.~\eqref{eq:FH3nosolvnewparams}. 
Following the standard Flory--Huggins procedure, we determine the critical conditions for phase separation by identifying compositions at which the free energy develops a critical point, characterized by the simultaneous vanishing of the second and third derivatives with respect to the independent volume fractions:
\begin{subequations}\label{eq:critical3com}
\begin{align}
\frac{1}{N_{3}(1-\phi_{1}-\phi_{2})}+\frac{1}{N_{1}\phi_{1}}-\xi_{1} &= 0, \label{eq:critical3com:a}\\
\frac{1}{N_{3}(1-\phi_{1}-\phi_{2})}+\frac{1}{N_{2}\phi_{2}}-\xi_{2} &= 0, \label{eq:critical3com:b}\\
\frac{1}{N_{3}(1-\phi_{1}-\phi_{2})^{2}}-\frac{1}{N_{1}\phi_{1}^{2}} &= 0, \label{eq:critical3com:c}\\
\frac{1}{N_{3}(1-\phi_{1}-\phi_{2})^{2}}-\frac{1}{N_{2}\phi_{2}^{2}} &= 0. \label{eq:critical3com:d}
\end{align}
\end{subequations}

Solving Eqs.~\eqref{eq:critical3com:a}--\eqref{eq:critical3com:d} yields closed-form expressions for the critical compositions $\phi_1^c$ and $\phi_2^c$, as well as the critical effective interaction parameters $\xi_1^c$ and $\xi_2^c$:
\begin{subequations}\label{eq:critical3comsol}
\begin{align}
\phi_1^c &=
\sqrt{\frac{N_{3}}{N_{1}}}
\left[
1+\sqrt{N_{3}}\!\left(\frac{1}{\sqrt{N_{1}}}+\frac{1}{\sqrt{N_{2}}}\right)
\right]^{-1},\\[6pt]
\phi_2^c &=
\sqrt{\frac{N_{3}}{N_{2}}}
\left[
1+\sqrt{N_{3}}\!\left(\frac{1}{\sqrt{N_{1}}}+\frac{1}{\sqrt{N_{2}}}\right)
\right]^{-1},\\[6pt]
\xi_1^c &=
\frac{\frac{1}{\sqrt{N_{3}}}+\frac{1}{\sqrt{N_{1}}}}{\sqrt{N_{3}}}
\left[
1+\sqrt{N_{3}}\!\left(\frac{1}{\sqrt{N_{1}}}+\frac{1}{\sqrt{N_{2}}}\right)
\right],\\[6pt]
\xi_2^c &=
\frac{\frac{1}{\sqrt{N_{3}}}+\frac{1}{\sqrt{N_{2}}}}{\sqrt{N_{3}}}
\left[
1+\sqrt{N_{3}}\!\left(\frac{1}{\sqrt{N_{1}}}+\frac{1}{\sqrt{N_{2}}}\right)
\right].
\end{align}
\end{subequations}
These critical quantities depend only on the effective degrees of polymerization of the three components and are independent of the microscopic interaction parameters, a consequence of the mean-field Flory–Huggins formulation and the neglect of explicit solvent degrees of freedom. 

The critical scaffold--client interaction strengths $\chi_{13}^c$ and $\chi_{23}^c$ follow directly from the definitions of the effective couplings,
\begin{equation}
\chi_{13}^c = \tfrac{1}{2}\!\left(\chi_{11}+\chi_{33}-\xi_1^c\right),
\label{eq:criticalhetero13}
\end{equation}
and
\begin{equation}
\chi_{23}^c = \tfrac{1}{2}\!\left(\chi_{22}+\chi_{33}-\xi_2^c\right).
\label{eq:criticalhetero23}
\end{equation}
These expressions show that the threshold for client recruitment depends explicitly on the scaffold-scaffold and client-client interactions.

The remaining critical parameter is the effective heterotypic interaction $\xi_3^c$ between the two scaffolds. 
At the spinodal surface, the Hessian matrix
\begin{equation}
H=
\begin{pmatrix}
\partial_{\phi_1}^2 f & \partial_{\phi_1\phi_2}^2 f\\
\partial_{\phi_2\phi_1}^2 f & \partial_{\phi_2}^2 f
\end{pmatrix}
\end{equation}
develops a zero eigenvalue, corresponding to $\det H=0$. 
Solving this condition at the critical composition yields
\begin{equation}
\xi_3^c =
\frac{1-
A N_3
\sqrt{
\frac{
\left(N_1\phi_1^c(A\xi_1^cN_3-1)-AN_3\right)
\left(N_2\phi_2^c(A\xi_2^cN_3-1)-AN_3\right)
}{
A^2 N_1 N_2 N_3^2 \phi_1^c\phi_2^c
}
}
}{A N_3},
\label{Eq:criticalheterotypic}
\end{equation}
where $A=1-\phi_1^c-\phi_2^c$ denotes the critical volume fraction of the client.

Using the definition of the effective heterotypic coupling, the critical scaffold--scaffold interaction strength can be written as
\begin{equation}
\chi_{12}^c
=
\xi_3^c-\chi_{33}+\chi_{13}^c+\chi_{23}^c ,
\label{eq:effective12}
\end{equation}
which simplifies to
\begin{equation}
\begin{aligned}
\chi_{12}^c
&=
\tfrac{1}{2}(\chi_{11}+\chi_{22})
-\tfrac{1}{2}\mathcal{K}^{(3)}, \\
\mathcal{K}^{(3)}
&=\xi_1^c+\xi_2^c-2\xi_3^c .
\end{aligned}
\label{eq:criticalparam3component}
\end{equation}

The quantity $\chi_{12}^c$ therefore defines the critical heterotypic scaffold--scaffold interaction strength at which the homogeneous mixture first becomes unstable. 
For $\chi_{12}>\chi_{12}^c$, the system undergoes associative liquid--liquid phase separation with positively sloped tie lines, whereas for $\chi_{12}<\chi_{12}^c$ the mixture remains homogeneous or exhibits segregative behavior. 
This criterion provides a clear thermodynamic boundary separating associative and aggregative phase separation in the presence of an explicit client.

\begin{figure*}[t]
\centering
\includegraphics[width=0.7\textwidth]{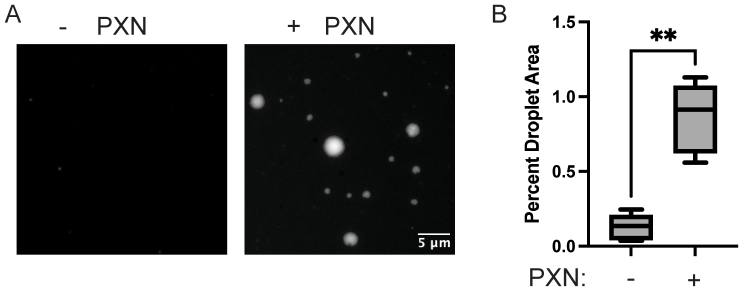}
\caption{
Effect of client addition on phase separation with two scaffolds (FAK and Cas) and one client (PXN).
FAK--Cas condensates were formed by combining GFP--pFAK, p130Cas, Nck, and N-WASP at 500~nM each in buffer containing 50~mM HEPES (pH~7.5), 65~mM NaCl, 2\% glycerol, and 1~mM DTT, either in the absence or presence of paxillin at 1~$\mu$M. Samples were imaged using epifluorescence microscopy after 30~minutes to allow the system to reach equilibrium.
A) Representative fluorescence images of GFP--FAK showing condensate formation. Scale bar, 5~$\mu$m.
B) Mean droplet area, defined as the fraction of the image occupied by the dense phase, quantified from five images and displayed as box plots. Statistical significance was assessed using Welch’s \emph{t}-test ($p < 0.0015$).
}
\label{fig:fig6}
\end{figure*}
For comparison, we also compute the critical heterotypic scaffold--scaffold interaction for the corresponding \emph{binary} (two-scaffold) incompressible Flory--Huggins model with solvent; the result is given in Eq.~\eqref{eq:criticalparam2component} (Appendix~A). 
A key consequence of PXN enrichment as an interacting component is that the location of the critical point shifts in interaction parameter space. Specifically, for fixed scaffold homotypic interactions and chain lengths, the critical heterotypic coupling required for associative liquid–liquid phase separation is reduced in the three-component scaffold–client model:
\begin{equation}
\begin{aligned}
\chi_{12}^{c}\big|_{\mathrm{binary}}
&>\chi_{12}^{c}\big|_{\mathrm{client}} \\
&=\chi_{12}^{c}
\;\text{(Eq.~\eqref{eq:criticalparam3component})}.
\end{aligned}
\label{eq:chi12c_inequality}
\end{equation}

In other words, introducing an explicit client generically \emph{facilitates} associative phase separation by lowering the minimum scaffold--scaffold heterotypic attraction required to destabilize the homogeneous state. At fixed microscopic scaffold--scaffold interactions $(\chi_{11}, \chi_{22}, \chi_{12})$, this renormalization shifts the system farther beyond criticality, expanding the two-phase region and enlarging the coexistence window in composition space. As a direct consequence, client recruitment is expected to lower the saturation concentration, $C_{\mathrm{sat}}$, required for condensate formation by stabilizing the dense phase at lower overall scaffold concentrations. This provides a thermodynamic mechanism by which client recruitment can regulate not only condensate \emph{composition} through selective partitioning, but also the \emph{distance to criticality} and the onset of phase separation itself, thereby modulating condensate sensitivity to perturbations. 

\subsection{In vitro experiments to test client effect on critical points}\label{exper}

To directly test whether clients facilitate associative phase separation, we returned to \emph{in vitro} experiments with reconstituted focal adhesion condensates.  We combined GFP-pFAK, phosphorylated Cas, Nck, and NWASP at $500~\text{nM}$ concentration, a concentration that did not readily form condensates. Including $1~\mu\text{M}$ PXN was sufficient to promote robust condensate formation (Fig.~\ref{fig:fig6} A,B). Thus, adding the PXN effectively shifts the FAK-CAS system beyond criticality. These experiments demonstrate that client recruitment is sufficient to change the phase behavior of a multi-scaffold system.

\section{Conclusion}\label{Sec:Conclusion}

In this work, we have shown that post-translational modification of scaffolds and recruitment of client proteins act as complementary and mechanistically distinct routes for regulating condensate phase behavior and composition. Using reconstituted experiments together with Flory–Huggins mean-field theory, we demonstrate that a client (PXN) selectively enriches condensates in one scaffold (FAK) while depleting another (Cas), and simultaneously facilitates its own incorporation. In our framework, phosphorylation modifies scaffold-scaffold interactions as well as the effective translational entropy, whereas client recruitment reshapes the interaction landscape by renormalizing scaffold--scaffold and scaffold--client couplings. These effects shift the location of the critical point in composition and interaction-parameter space, thereby lowering the threshold for phase separation. As a result, client recruitment, mediated by homotypic client interactions, drives the system further from criticality, enlarges the two-phase coexistence region, and enhances selective enrichment in the dense phase, as demonstrated in our reconstituted experiments.

The present theory is formulated as an equilibrium, mean-field description of multicomponent phase separation. As such, it captures phase coexistence, critical behavior, and composition trends driven by effective interactions, but it does not explicitly account for saturable binding, finite valency, network topology, or spatial heterogeneity within condensates. An important extension would be to examine these effects using explicit sticker-spacer models, which naturally incorporate valency constraints, sequence architecture, and network connectivity~\cite{choi2020physical,lin2023heterotypic, choi2019lassi, grandpre2023linker,zhang2021decoding, akram2025biomolecular}.  Moreover, phosphorylation is treated here as shifting the system between distinct equilibrium states. In living cells, however, phosphorylation and dephosphorylation are often maintained far from equilibrium by continuous ATP consumption and enzymatic turnover. Extending this framework to include nonequilibrium effects, such as time-dependent modification cycles, ATP-driven turnover, or reaction--diffusion coupling between phase separation and biochemical reactions, represents a natural direction for future work~\cite{wurtz2018chemically,weber2019physics, soding2020mechanisms, glotzer1995reaction}. 

An important extension of this work would be to examine how client--scaffold interactions couple condensate thermodynamics to membrane association and wetting.
Biomolecular condensates frequently interact with membranes through specific tethers and can form membrane-bound two-dimensional condensates that coexist with or nucleate three-dimensional condensates, thereby linking bulk phase behavior to membrane properties~\cite{zhao2021thermodynamics,rouches2021surface,sun2022kinetic,yu2024membrane, bagheri2026membrane, kato2025wetting}. 
In the FAK–Cas system, both scaffolds associate with the plasma membrane, with FAK binding directly to the membrane~\cite{litschel2024membrane, acebron2020structural}. 
Client-mediated renormalization of scaffold interactions, as identified here, is therefore expected to influence not only condensate composition but also membrane wetting, interfacial tension, and the stability of membrane-bound condensates. 
More broadly, extending the present framework to include additional clients would enable systematic investigation of how multiple recruitment pathways cooperatively or competitively reshape scaffold-scaffold interactions, dilute-phase concentrations, and oligomeric states, as observed in multicomponent condensates formed by RNA-binding proteins and signaling assemblies~\cite{ge2025fus,he2023phase, sundaravadivelu2024sequence}.

Overall, our results provide a unified thermodynamic framework for understanding how phosphorylation and client recruitment jointly regulate the composition, stability, and critical properties of multicomponent biomolecular condensates. By clarifying how effective interactions and mixing entropy can be independently tuned, this work bridges reconstituted experiments and theoretical models, and offers general design principles for how cells dynamically control condensate function through multicomponent regulation.

\section{Acknowledgments}
We thank Qiwei Yu for useful discussions and for sharing code. T.G. acknowledges support from, and stimulating discussions at, the National Institute for Theory and Mathematics in Biology (NITMB). L.C. and N.E.L. acknowledge support from the Air Force Office of Scientific Research under Grant No. FA9550-22-1-0207. This work was supported in part by National Science Foundation Grant No. PHY-2309135 and by the Gordon and Betty Moore Foundation Grant No. 2919.02 to the Kavli Institute for Theoretical Physics (KITP). \newline

\appendix
\renewcommand{\thefigure}{S.\arabic{figure}}
\setcounter{figure}{0} 
\section*{Appendix A: Critical interaction parameter for two component system}
\label{App:A}
\setcounter{equation}{0}
\renewcommand{\theequation}{A.\arabic{equation}}
In this subsection we calculate the critical interaction parameter for associative phase separation for two interacting scaffold molecules in absence of any explicit solvent.

The critical interaction parameter for two scaffold molecules interacting with each other for which the phase separation occurs in the absence of any solvent. Substituting $\phi_{2} = 1- \phi_{1}$ into Eq.~\eqref{eq:FH3}, we find 
\begin{widetext}
\begin{equation}
   f_{-s} = \frac{\phi_{1}}{\text{N}_{1}} \log \phi_{1} + \frac{(1-\phi_{1})}{\text{N}_{2}} \log (1-\phi_{1})-\frac{1}{2} (\chi_{11}+\chi_{22}-2 \chi_{12}) \phi_{1}^2 +(\chi_{22}-\chi_{12})\phi_{1} -\frac{1}{2}\chi_{22}
   \label{eq:fh2component}
\end{equation}
\end{widetext}
The subscript ``$-s$" stands for absence of solvent.
To obtain the critical parameters, we set $\partial^{2}f/\partial \phi_{1}^2=0$ and $\partial^{3}f/\partial \phi_{1}^3=0$, which leads to
\begin{equation}
    \frac{1}{N_{2} (1-\phi_{1})}+\frac{1}{N_{1}\phi_{1}}-\tilde{\chi} = 0 
    \label{eq:criticalInteractionval}
\end{equation}
and
\begin{equation}
\frac{1}{N_{2}(1-\phi_1)^2}-\frac{1}{N_{1} \phi_{1}^2} = 0 
\end{equation}
where $\tilde{\chi} = \chi_{11} + \chi_{22} -2 \chi_{12}$ in Eq.~\eqref{eq:criticalInteractionval}. 
Solving these equations result 
\begin{equation}
\tilde{\chi}^c = \left(\frac{1}{\sqrt{N_{1}}} +\frac{1}{\sqrt{N_{2}}}\right)^2
\end{equation}
\\
From there, we can obtain the effective heterotypic interaction, 
\begin{equation}
\chi_{12}^{c} = \frac{1}{2} \left(\chi_{11} + \chi_{22}\right) - \frac{1}{2}\tilde{\chi}^c,
\end{equation}
which further reduces to
\begin{equation}
\chi_{12}^{c} = \frac{1}{2} \left(\chi_{11} + \chi_{22}\right) - \frac{1}{2}\left(\frac{1}{\sqrt{N_{1}}} +\frac{1}{\sqrt{N_{2}}}\right)^2.
\label{eq:criticalparam2component}
\end{equation}

Heterotypic interaction parameter larger than $\chi_{12}^{c}$ indicates the associative LLPS.

\end{document}